\documentclass[aip, jmp,amsmath,amssymb,reprint]{revtex4-1}

\usepackage{graphicx}
\usepackage{bm}
\usepackage{subfig}
\usepackage{epstopdf}
\usepackage{natbib}
\usepackage{float}

\draft 

\begin{document}


\title{A high-precision mechanical absolute-rotation sensor} 



\author{Krishna Venkateswara}
\email[]{kvenk@uw.edu}
\affiliation{Center for Experimental and Nuclear Physics and Astrophysics, University of Washington, Seattle, Washington, 98195, USA}

\author{Charles A. Hagedorn}

\author{Matthew D. Turner}

\author{Trevor Arp}

\author{Jens H. Gundlach}


\date{\today}

\begin{abstract}
We have developed a mechanical absolute-rotation sensor capable of resolving ground rotation angle of less than 1 nrad$/\sqrt{\text{Hz}}$ above $30$ mHz and 0.2 nrad$/\sqrt{\text{Hz}}$ above $100$ mHz about a single horizontal axis. The device consists of a meter-scale beam balance, suspended by a pair of flexures, with a resonance frequency of 10.8 mHz. The center of mass is located 3 $\mu$m above the pivot, giving an excellent horizontal displacement rejection of better than $3\times10^{-5}$ rad/m. The angle of the beam is read out optically using a high-sensitivity autocollimator. We have also built a tiltmeter with better than 1 nrad$/\sqrt{\text{Hz}}$ sensitivity above 30 mHz. Co-located measurements using the two instruments allowed us to distinguish between background rotation signal at low frequencies and intrinsic instrument noise. The rotation sensor is useful for rotational seismology and for rejecting background rotation signal from seismometers in experiments demanding high levels of seismic isolation, such as Advanced LIGO.  
\end{abstract}


\maketitle 


\section{Introduction}

It is well known that horizontal seismometers are sensitive to ground rotation at low frequencies.\cite{seism1,seism2, tiltfree} Indeed, due to the equivalence principle, conventional seismometers and tiltmeters cannot distinguish between horizontal acceleration and rotation of the ground. While this is a problem in precision seismology, it is especially problematic for seismic isolation in next-generation gravitational-wave detectors, such as Advanced Laser Interferometer Gravitational-wave Observatory (aLIGO), where it is believed that rotation noise may limit improvements in low-frequency isolation.\cite{reqs}

\begin{figure}[!ht]
\includegraphics{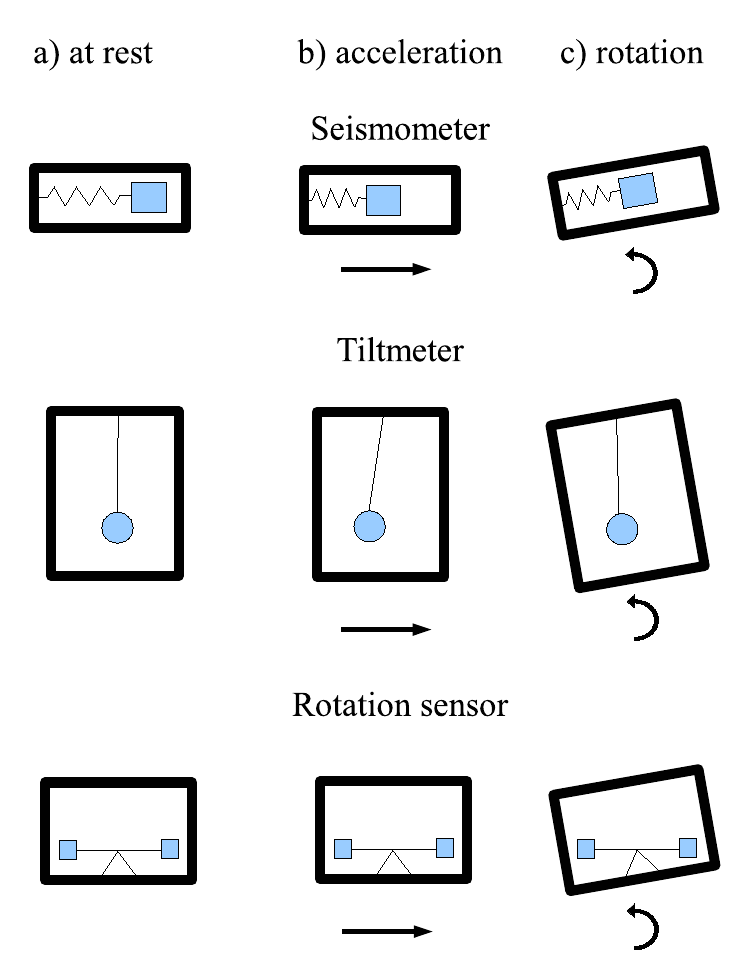}
\caption{\label{tiltaccel}Comparison between a horizontal seismometer, a tiltmeter, and our rotation sensor under the influence of slow horizontal acceleration and rotation. In the first two instruments, an observer inside the box cannot distinguish between the displacement of the bob due to rotation and acceleration.}
\end{figure}

Conventional horizontal seismometers can be idealized as horizontal spring-mass systems whose horizontal displacement is sensed relative to the housing. Similarly, conventional tiltmeters or inclinometers can be idealized as vertical pendulums whose horizontal displacement is sensed relative to the housing. They are schematically represented in Fig.~\ref{tiltaccel}. By `tilt', we henceforth refer to the angular deflection of a simple pendulum with respect to its suspension platform or enclosure.

From the diagram of the horizontal seismometer, a periodic rotation of $\theta$ at angular frequency $\omega$ will look equivalent to an acceleration of $g\theta$ or a displacement of $-g\theta/\omega^2$. In other words, the displacement in response to a unit rotation for the horizontal seismometer is
\begin{equation}
\text{response to unit rotation}=-\frac{g}{\omega^2}.
\label{eq0}
\end{equation}
Due to the $\omega^2$ in the denominator, response to rotation dominates at low frequencies (typically $f<30$ mHz). Similarly for a pendulum tiltmeter, the displacement of the bob due to a 
rotation is indistinguishable from that due to a horizontal acceleration at frequencies well below the resonance frequency. Consequently, the rotation in response to a unit displacement is given by the inverse of the right-hand side of Eq.~(\ref{eq0}). Thus, typically, a tiltmeters output is dominated by acceleration at frequencies greater than $\sim80$ mHz.

An important limitation of the active-control system for seismic isolation in aLIGO is this inability of horizontal seismometers to distinguish between ground rotation and horizontal acceleration at low frequencies ($10$ to $500$ mHz\cite{reqs}).  Slow ground rotation, such as that induced by wind, would be misinterpreted as large ground displacements, which could cause a large correction to be applied at the low-frequency active-isolation stages. This large displacement can produce noise in the gravitational-wave signal band through frequency up-conversion mechanisms, non-linearities, and cross-couplings. This problem can be addressed by measuring the absolute ground rotation and removing it from the seismometer channels before it has a chance to enter the isolation system.

Our rotation sensor design may also be important to the field of rotational seismology.\cite{tiltmeter} Large ring-laser gyroscopes are the traditional rotation sensors used in this field.\cite{gyro,gyro2} They are traditionally run horizontally to sense rotation about the vertical axis, but can be mounted vertically to sense rotation about a horizontal axis.\cite{gyro3,belfi} Our design offers a simpler and more compact alternative. For comparison, it has roughly an order of magnitude better angle sensitivity than the horizontal-axis ring-laser gyroscope described in Belfi \textit{et al.}\cite{belfi} between 10 to 100 mHz. In this frequency band, our sensor has comparable sensitivity to the angle sensitivity of C-II: a vertical-axis meter-scale monolithic laser ring gyro\cite{gyro,gyro2} and its sensitivity is surpassed by roughly an order of magnitude by the horizontal-axis 3.5 meter-square G-0 ring-laser gyro.\cite{gyro2} 

\section{Principle}
\begin{figure}
\includegraphics{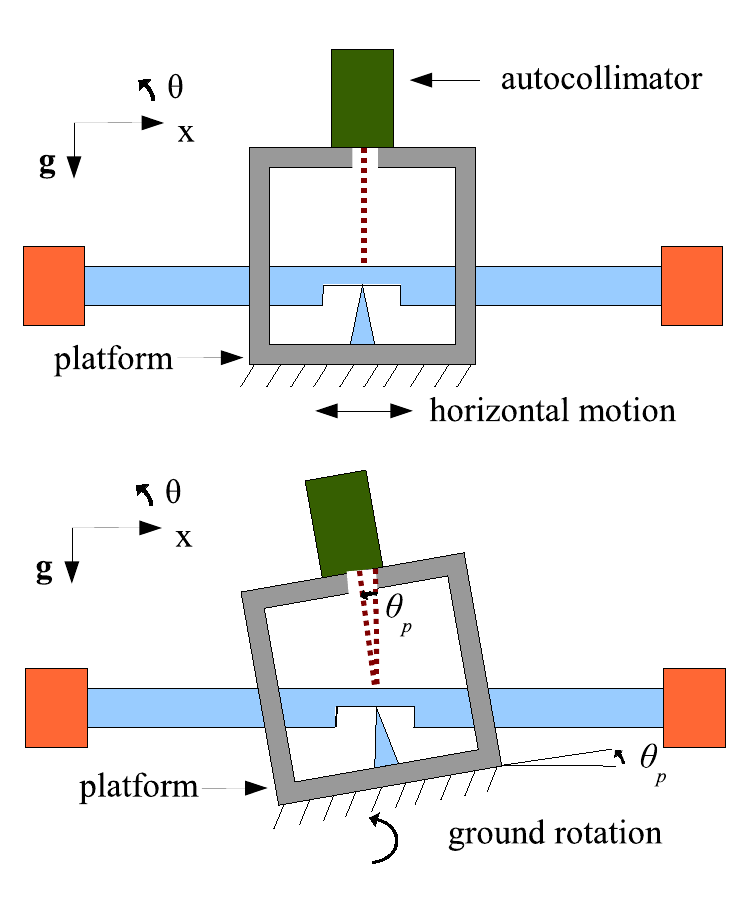}
\caption{\label{Schematic} Schematic showing the principle of the rotation sensor. At frequencies above the resonance of the balance, it stays inertial as the platform rotates. Thus the autocollimator measures the platform rotation.}
\end{figure}

Our rotation sensor is a low-frequency beam balance whose angle with respect to the platform is measured using an autocollimator. Above the resonance frequency, the beam balance stays inertial as the platform rotates around it. Thus, the autocollimator measures the platform rotation, as shown in Fig.~\ref{Schematic}. To decouple the influence of rotation and translation, the center of mass (COM) of the balance is located as close to the axis of rotation as possible. The relevant parameters of the balance are listed in Table \ref{paratab}.

\begin{table}
\caption{\label{paratab}Parameters of the balance}
\begin{ruledtabular}
\begin{tabular}{cc}
Parameter & Value \\
\hline
Total mass M & $4.0$ kg\\
Moment of inertia I & $0.51$ kg $\text{m}^2$\\
Resonance frequency & $10.8$ mHz\\
Center of mass offset $\delta$ & $-3\pm1$ $\mu$m\\
Flexure angular stiffness $\kappa_e$ & $2.5\times10^{-3}$ N m/rad\\
Flexure thickness & $(25\pm2)\times10^{-6}$ m\\
Quality factor Q & $5\times10^3$\\
\end{tabular}
\end{ruledtabular}
\end{table}

\subsection{Equation of motion}
\label{subsec:eqnmotion}
To understand the dynamics of the system, we can write down the equations of motion in an inertial frame aligned with gravitational vertical. Let $\theta$ be the angle between the beam and the inertial frame's horizontal plane and let $\theta_p$ be the platform rotation angle with respect to the inertial frame's horizontal plane. The equation of motion for rotation of the balance about the pivot axis in the presence of external torques (assuming $\theta\ll1$) is
\begin{equation}
I\ddot{\theta} + \kappa_e (1+i\phi)(\theta-\theta_p)+Mg\theta\delta+M\ddot{x_p}\delta=\tau_{ext},
\label{eq1}
\end{equation}
where $\delta$ is the vertical distance from the COM and the pivot (positive sign if the COM is below the pivot), $\kappa_e$ is the stiffness of the flexure, $M$ is the total suspended mass of the balance, $I$ is its moment of inertia, $\phi$ is the loss factor of the flexure material ($\phi=1/Q$), and $x_p$ is the horizontal displacement of the platform. The external torque $\tau_{ext}$ includes torques from all sources other than acceleration or rotation, such as Brownian motion ($k_BT$ noise)\cite{thermal}, temperature gradients, or magnetic fields. Accelerations and rotations in other degrees of freedom are ignored in this simple model as they couple only through higher-order effects.

To establish the transfer function for the autocollimator output in response to a platform rotation, external torque and acceleration can be set to zero. Substituting $\theta = \hat{\theta} e^{i\omega t}$ and $\theta_p = \hat{\theta}_p e^{i\omega t}$ and solving for $\hat{\theta}$ gives
\begin{equation*}
\hat{\theta} = -\hat{\theta}_p\frac{\kappa_e(1+i\phi)}{I\omega^2-\kappa_e(1+i\phi)-Mg\delta}.
\end{equation*}
The quantity measured by the autocollimator is
\begin{equation}
\hat{\theta}_a=\hat{\theta}-\hat{\theta}_p = -\hat{\theta}_p\frac{I\omega^2-Mg\delta}{I\omega^2-\kappa_e(1+i\phi)-Mg\delta}
\label{autocoleq}
\end{equation}
and the transfer function is
\begin{equation}
\frac{\hat{\theta}-\hat{\theta}_p}{\hat{\theta}_p} = -\frac{\omega^2-\frac{Mg\delta}{I}}{\omega^2-\frac{\kappa_e(1+i\phi)}{I}-\frac{Mg\delta}{I}}.
\label{tiltresp}
\end{equation}

The magnitude of the transfer function can be rearranged into the usual form for an oscillator as
\begin{equation}
\left|\frac{\hat{\theta}-\hat{\theta}_p}{\hat{\theta}_p}\right| = \frac{|\omega^2-\omega_g^2|}{\sqrt{(\omega^2-\omega_0^2)^2+\frac{\omega_0^4}{Q'^2}}},
\label{tiltresp2}
\end{equation}
where $\omega_0^2=\frac{\kappa_e+Mg\delta}{I}$, $\omega_g^2=\frac{Mg\delta}{I}$ and $Q'=Q\frac{\kappa_e+Mg\delta}{\kappa_e}$. The new quality factor $Q'$ can be either enhanced or reduced depending upon the sign of $\delta$. The transfer function has a zero at $\omega_g$  and a pole at $\omega_0$. At frequencies much less than both $\omega_g$ and $\omega_0$,
\begin{equation}
\left|\frac{\hat{\theta}-\hat{\theta}_p}{\hat{\theta}_p}\right| \approx \left|\frac{\omega_g^2}{\omega_0^2}\right|.
\label{eq6}
\end{equation}

\begin{figure}
\includegraphics{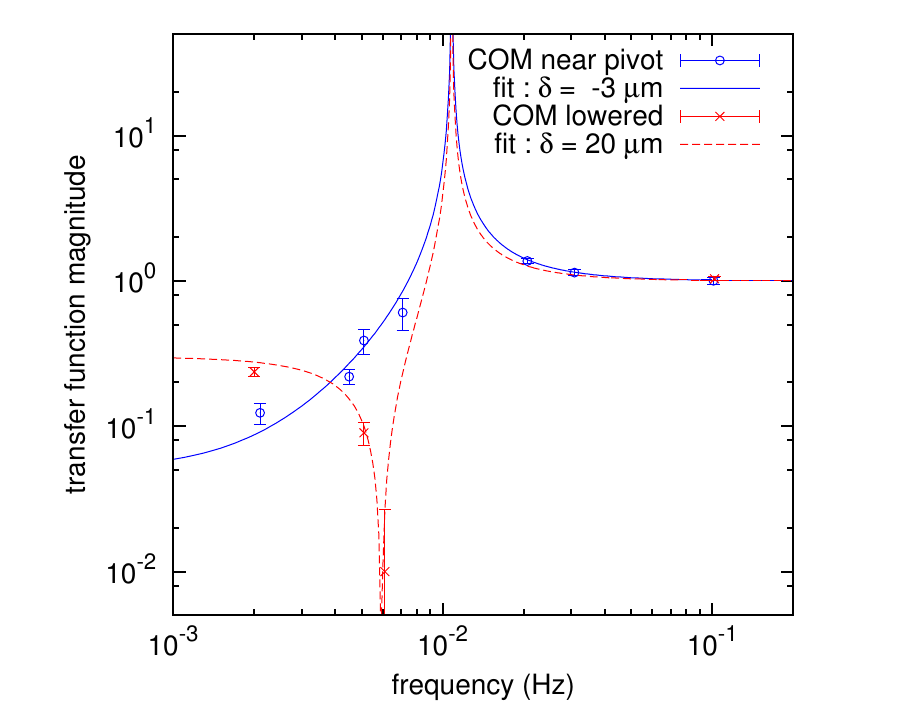}
\caption{\label{TransFunc} Magnitudes of the transfer function between platform rotation and autocollimator output for two configurations of the balance with different COM. The measurement and a fit with free parameter $\delta$ are shown for each configuration. A positive $\delta$ results in a real zero in the transfer function whereas a negative $\delta$ yields an imaginary zero.}
\end{figure}

Fig.~\ref{TransFunc} shows the measured transfer function for a positive and a negative $\delta$. The low frequency response is given by Eq.~(\ref{eq6}), which is strongly sensitive to $\delta$. Based on the sign of the gravitational spring stiffness $Mg\delta$ there are three possibilities:
\begin{enumerate}
\item When $Mg\delta$ is much larger than $\kappa_e$, the balance behaves as a simple pendulum, giving a low frequency response of $1$. 
\item When $Mg\delta$ is close to zero, the balance behaves as a spring-mass system, giving a low frequency response much less than $1$. 
\item When $Mg\delta$ is $\sim-\kappa_e$, the low frequency response is enhanced. This principle was exploited by building a second balance (Sec.~\ref{rotdata}) with a negative and large $\delta$, which made it very sensitive to tilt.
\end{enumerate}

Thus, a simple way of minimizing $|\delta|$ is to minimize the low-frequency rotation response of the balance. This measurement technique provides a method for minimizing $|\delta|$ far more precisely than would otherwise be possible.

Similarly, we can calculate the transfer function for the autocollimator output in response to a platform acceleration. Setting external torque and platform rotation to zero gives
\begin{equation}
\frac{\hat{\theta}}{\hat{x}_p} = -\frac{M\omega^2\delta}{I\omega^2-\kappa_e(1+i\phi)-Mg\delta}.
\label{acceq}
\end{equation}
Rearranging terms as in Eq.~(\ref{tiltresp2}) yields
\begin{equation*}
\text{response to unit displacement}=\frac{M\omega^2\delta}{I(\omega^2-\omega_0^2-i\frac{\omega_0^2}{Q'})},
\end{equation*}
and for $\omega\gg\omega_0$,
\begin{equation}
\text{response to unit displacement} \approx \frac{M\delta}{I}.
\label{respeq2}
\end{equation}
For our balance this is less than $3\times10^{-5}$ rad$/$m. This ratio is equivalent to that of a 33-km-long simple pendulum.

\subsection{Torques acting on the balance}
A useful way of understanding the noise sources on the balance is to calculate the net torque acting on the system. The torque measured in the non-inertial lab frame can be written as
\begin{equation}
\tau_{lab} = I\ddot{\theta_a}+I\omega_0^2(1+i\phi')\theta_a.
\label{torqeq}
\end{equation}
Using Eqs.~(\ref{eq1}) and (\ref{autocoleq}) we can rewrite this as
\begin{equation*}
\tau_{lab} = \tau_{ext}-I\ddot{\theta}_p-Mg\delta\theta_p-M\delta\ddot{x},
\end{equation*}
and converting to the frequency domain yields
\begin{equation}
\hat{\tau}_{lab} = \hat{\tau}_{ext}+I\omega^2\hat{\theta}_p-M\delta(g\hat{\theta}_p-\hat{x}\omega^2).
\label{eq9}
\end{equation}
As mentioned previously, the first term includes torques from all non-acceleration sources. The second term in Eq.~(\ref{eq9}) contains the signal we are interested in measuring. This can also be interpreted as a pseudo-torque. The third term is the torque due to accelerations that couple through the gravitational spring.

\subsection{The Instrument}

\begin{figure}
\includegraphics{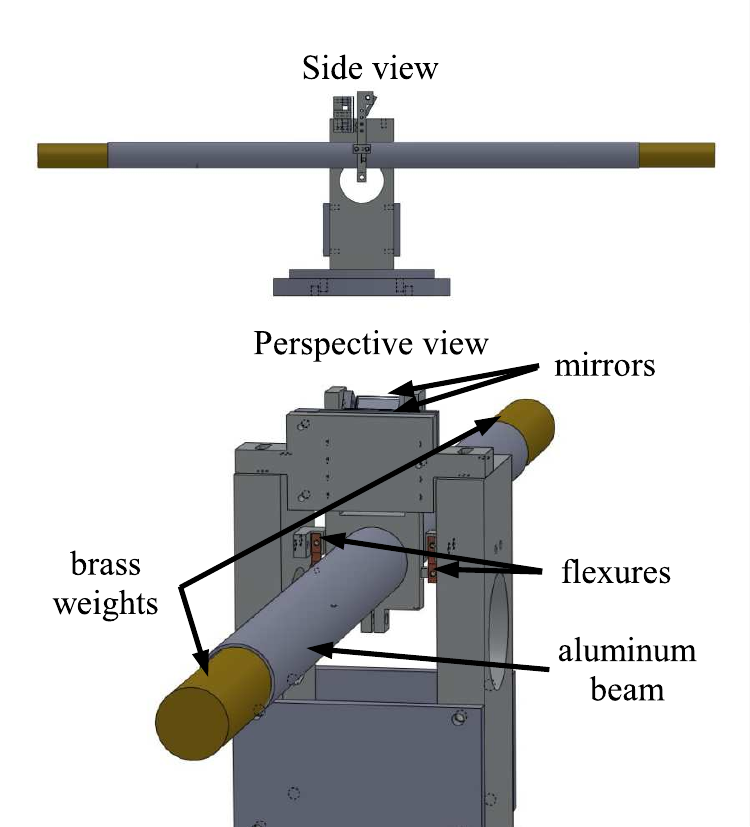}
\caption{\label{Model} Drawing of the rotation sensor showing the beam balance suspended by a pair of flexures. Mirrors are located at the center for readout of the angular position of the balance.}
\end{figure}

\begin{figure}
\includegraphics{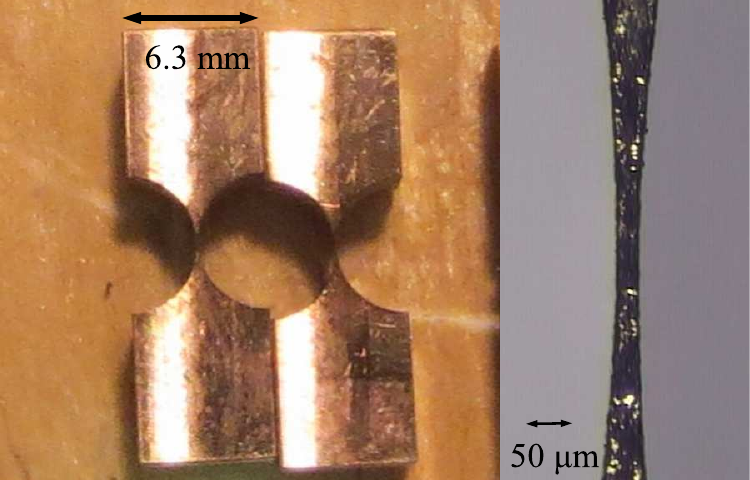}
\caption{\label{flex} A pair of beryllium-copper flexures are shown on the left. The right side shows an image of the flexure as  viewed through a microscope.}
\end{figure}

The balance beam consists of an aluminum tube with brass weights at the ends. The tube has a diameter of 38 mm, wall thickness of 1.6 mm, and a length of 0.76 m. The brass weights are solid cylinders of diameter 34 mm and length of 0.23 m, and are fitted into the aluminum tube bringing the end to end length of the balance to 0.97 m. The balance is suspended by two flexures machined from beryllium-copper alloy, shown in Figs.~\ref{Model} and \ref{flex}. They are notch type flexures, with a thickness of $25\pm2$ $\mu$m, radius of 3.2 mm, and width of 6.3 mm. Following standard procedures for achieving optimum tensile strength, the flexures were age-hardened at $300^{\circ}$ C for about 2 hours.

\begin{figure}
\includegraphics{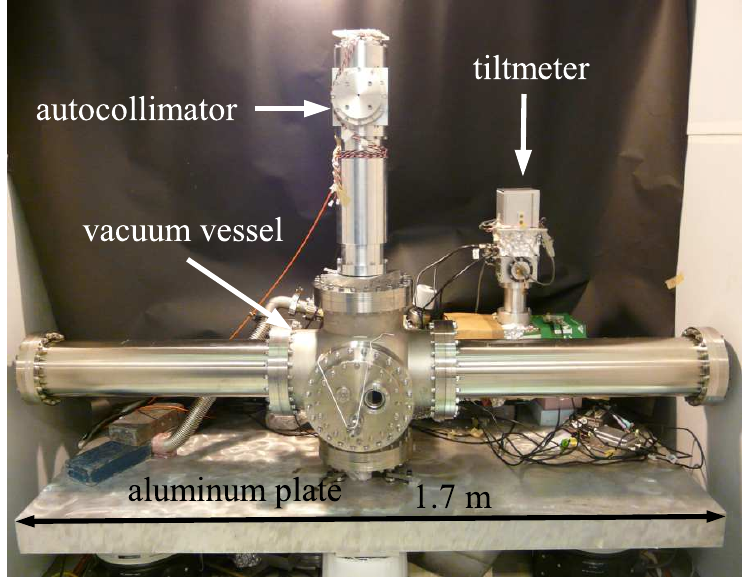}
\caption{\label{can} University of Washington rotation sensor. Thermal insulation was removed for this picture.}
\end{figure}

We built a vacuum chamber for the balance using stainless steel CF vacuum components (see Fig.~\ref{can}). The chamber consists of a 6-way cross with two tubes flanged to it. A turbo vacuum pump backed by a diaphragm pump is connected through a long bellows to the 6-way cross. The side-port contains a mechanical-manipulator bellows to adjust the horizontal COM while the balance is under vacuum. A viewport on top allows optical access for an autocollimator to measure the angle of the balance.

The instrument is located in a temperature-controlled ($\pm0.1^{\circ}$ C) underground experimental hall and is surrounded by an extruded-foam enclosure. The instrument is installed on a $0.1$-m-thick 600-kg aluminum plate bolted to three stainless steel feet. The feet sit on a 0.3 m-thick concrete platform that rests on a deep foundation. We installed 6.3-mm-thick aluminum thermal shielding tubes surrounding each balance-arm to provide thermal isolation and reduce thermal gradients. These aluminum tubes are located inside the vacuum vessel and are thermally decoupled from the vacuum chamber and from the balance, but are connected to each other through thick aluminum heat links.

The platform-angle-to-autocollimator-output transfer function was measured by applying a periodic horizontal force to the aluminum platform on which the balance sits. This force deflected the platform rotationally ($\pm0.3$ $\mu$rad) and induced horizontal translations ($\pm0.3$ $\mu$m). The periodic force was generated by a spring attached to a $20$-mm-diameter crankshaft driven by a stepper motor. The spring was connected to the platform through a $2.0$ m cable which reduced the variation in the direction of the applied force. The circular motion of the motor was thus converted to a linear force on the platform. Due to the small $\delta$ during measurements, the response of the balance to displacements of the platform were many orders of magnitude smaller than the rotation response. Hence the platform displacement did not affect the transfer function measurement. Each point in Fig.~\ref{TransFunc} represents a single measurement which consisted of driving the platform at a single frequency for a few thousand seconds and measuring the autocollimator output.

\subsection{Autocollimator}
\begin{figure}
\includegraphics{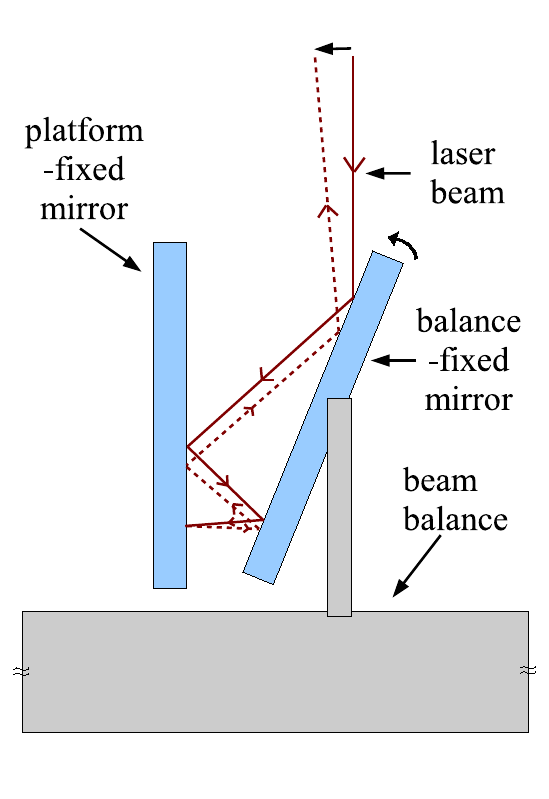}
\caption{\label{fourb} Schematic drawing  of the four reflections of the light beam on the target mirror.}
\end{figure}

We used a multi-slit autocollimator, built by our group, that measures the angular deflection of a light beam reflected off a mirror mounted on the balance.\cite{trevor} The autocollimator uses a fiber-coupled LED and a multi-slit grating as the light source and a sensitive CCD sensor for readout. Sensitivity was enhanced by employing multiple reflections of the light beam. Our setup used four reflections (see Fig.~\ref{fourb}) giving a dynamic range of 5 mrad.

The autocollimator is designed for use with a partially reflecting reference mirror in the beam path. Subtracting the reference pattern allows a rejection of common-mode noise such as low-frequency distortion of the vacuum can or the autocollimator body. In practice, we found such common-mode noise to be negligible compared with the seismic background. However, this option allows referencing to a surface other than the platform to which the balance is bolted. This would prove useful if we were to read the rotation of a surface that could not support the weight of the entire balance. The reference pattern readout is also a measure of the intrinsic autocollimator noise.

\subsection{Calibration}
The autocollimator was calibrated using one ordinary mirror and one partially reflecting mirror held at a fixed, known angle.\cite{trevor} In addition the instrument was calibrated using the gravitational torque from a pair of lead bricks on a turntable located $0.8$ m from the vacuum vessel. The computed gravitational torque agreed with the measured torque to within the $10$ percent systematic uncertainty. 

\section{Rotation data}
\label{rotdata}
\begin{figure}
\includegraphics{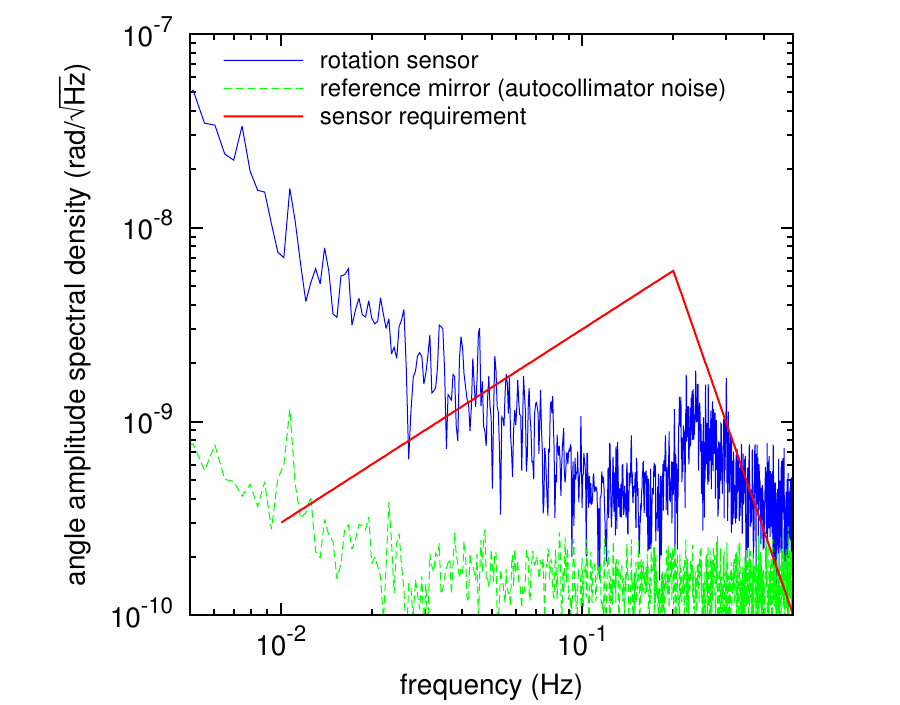}
\caption{\label{tiltdata} Typical platform rotation spectral density as measured by the rotation sensor during quiet conditions. Also shown are the autocollimator sensitivity and the rotation sensor requirement to potentially improve aLIGO.\cite{reqs}}
\end{figure}

The rotation sensor output was recorded for several months. Typically, the rotation noise was smallest from 22:00 hrs to 04:00 hrs. The rotation in the 10-to 100-mHz frequency range was dominated by local weather, especially wind. To understand the spectral distribution, the raw time series of the autocollimator output is Fourier transformed to an amplitude spectral density.  To obtain the platform rotation, we divide the autocollimator output by the measured transfer function between them (Eq.~(\ref{tiltresp})). Fig.~\ref{tiltdata} shows the spectral density of the data from a 4000-s measurement during a quiet period. The spectral density shown is typical for most nights with low wind speeds.

\begin{figure}
\includegraphics{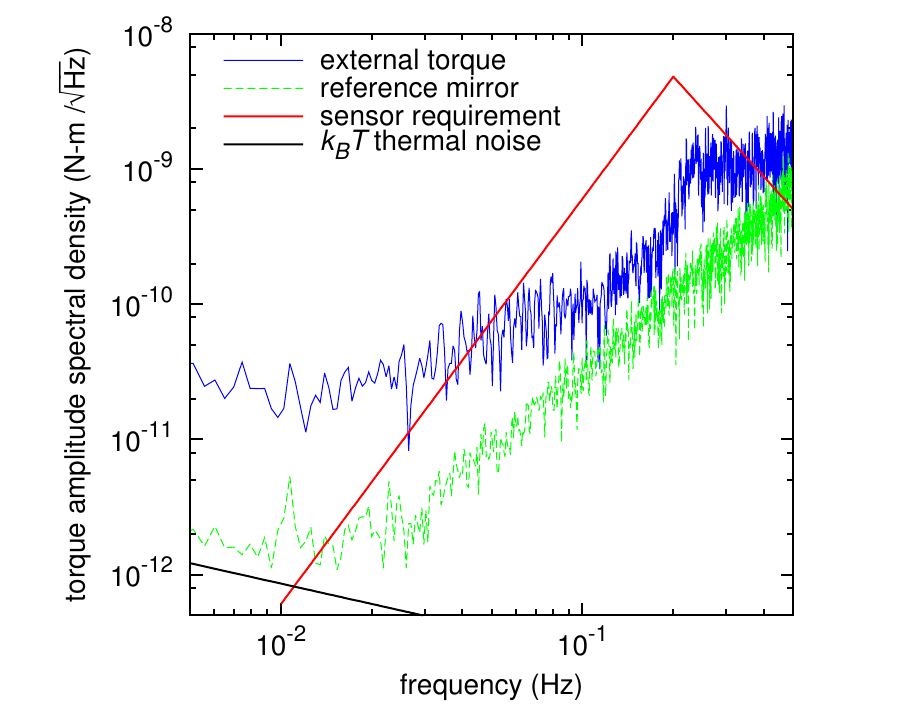}
\caption{\label{Torque} Torque sensitivity of the rotation sensor. Also shown are the sensor requirements, autocollimator limit and thermal noise limit\cite{thermal} in torque units.}
\end{figure}

Fig.~\ref{Torque} shows a plot of the torque spectral density as a function of frequency calculated using Eq.~(\ref{torqeq}) for the same data set as above. Also shown are the thermal or Brownian motion noise torque\cite{thermal} and the rotation sensor requirement,\cite{reqs} interpreted in torque units for our balance. 

\begin{figure}
\includegraphics{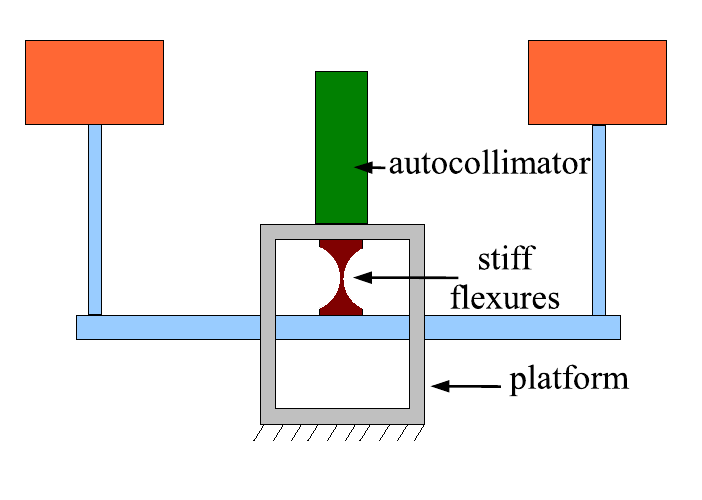}
\caption{\label{newtilt}Schematic of the tiltmeter. It is suspended by two stiff flexures. The COM is located 78 mm above the pivot. The angle of the beam is read-out using an autocollimator.}
\end{figure}

Separating the background rotation signal from the instrument noise requires a second instrument of similar or better sensitivity. Since the rotation background at frequencies above 50 mHz was sufficiently small during quiet conditions, we focused on measuring the ground rotation near the 10-mHz limit of the required bandwidth. To build a sensitive tiltmeter at these frequencies we exploited the principle of the amplification of rotation using a negative $\delta$, described in Sec.~\ref{subsec:eqnmotion}. We constructed the tiltmeter from an aluminum plate and brass weights suspended by two stiff flexures (Fig.~\ref{newtilt}). This balance had its COM far above the pivot ($\delta=-78$ mm) resulting in a negative and large gravitational stiffness. Thus, an autocollimator mounted to the frame measured an amplified rotation signal at frequencies below resonance of the tiltmeter.

This tiltmeter was placed on the same aluminum platform as the rotation sensor. To check the calibration and sensitivity of this instrument, we temporarily modified the rotation sensor to run in tiltmeter mode by changing the COM to be $1$ mm below the pivot. The difference between the tilt measurements by the two sensors was smaller than $3$ nrad/$\sqrt{\text{Hz}}$ at 10 mHz and less than $1$ nrad/$\sqrt{\text{Hz}}$ at 30 mHz. 

\begin{figure}
\includegraphics{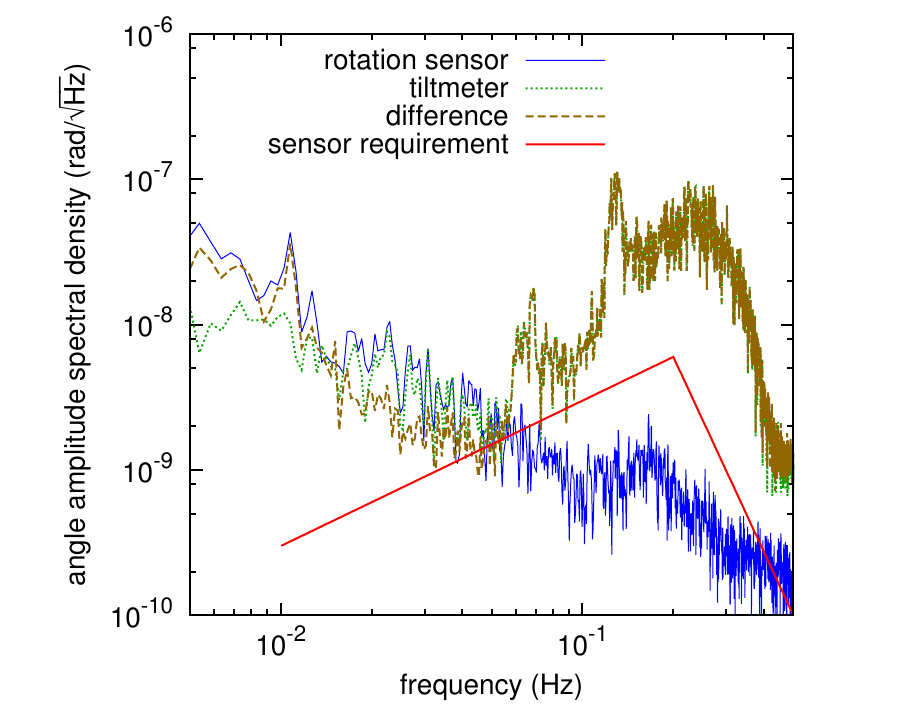}
\caption{Platform rotation spectral density as measured by the rotation sensor and the tiltmeter during the day, and the difference between the two. The broad peak in the tiltmeter 	spectral density is due to the microseismic peak. The roll-off in the tiltmeter output at 0.3 Hz is due to an anti-aliasing filter.}
\label{corr}
\end{figure}

\begin{figure}
\includegraphics{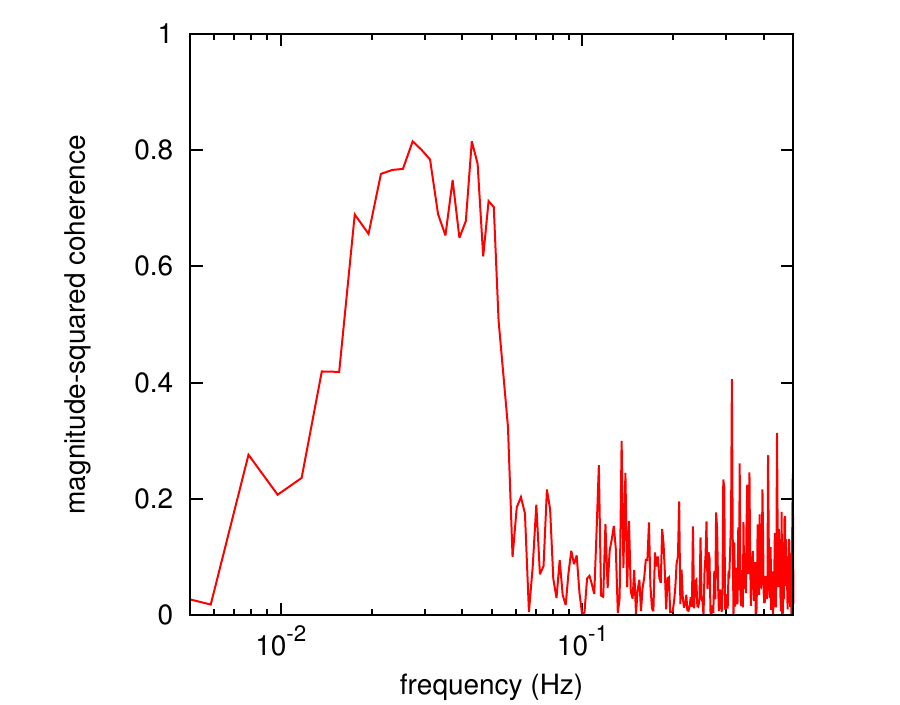}
\caption{Magnitude-squared coherence between the rotation sensor and the tiltmeter for the same data set as in Fig.~\ref{corr}.}
\label{msc}
\end{figure}

Fig.~\ref{corr} shows the angle spectral density from a 4000-s dataset recorded during the day with both instruments. The angle measured by the rotation sensor and the ordinary tiltmeter agree well in the 15-to 60-mHz frequency range. Above that, the angle measured by the tiltmeter is dominated by acceleration, which the rotation sensor rejects efficiently. Fig.~\ref{msc} shows the mean-squared coherence of the two data sets as a function of frequency. Subtracting the two datasets gives a factor of 2 to 3 suppression of rotation noise in the rotation sensor in the 15-to 60-mHz range. The small $\delta$ of the rotation sensor balance implies that the microseismic acceleration affecting the tiltmeter in Fig.~\ref{msc} near $0.2$ Hz would produce an angle noise of less than 10 picorad/$\sqrt{\text{Hz}}$. This is confirmed by the coherence plot which shows poor coherence at the microseismic frequencies. However, the residual noise after subtraction is worse compared to the noise from the quiet-condition data. We believe this excess noise arises from high-frequency vibration, discussed in the next section. During quiet conditions at night, when background rotations were much smaller, the two instruments showed reduced correlation, indicating that at least one instrument was limited by intrinsic noise. 

\section{Noise sources}
There are various environmental influences that can produce noise in the rotation sensor. Due to the low-pass filtering of torques  by the balance and the nature of background influences (such as weather), the noise requirement at 10 mHz is most challenging. Other than the autocollimator noise, all other noise sources are practically insignificant above 50 mHz. 

\begin{figure}
\includegraphics{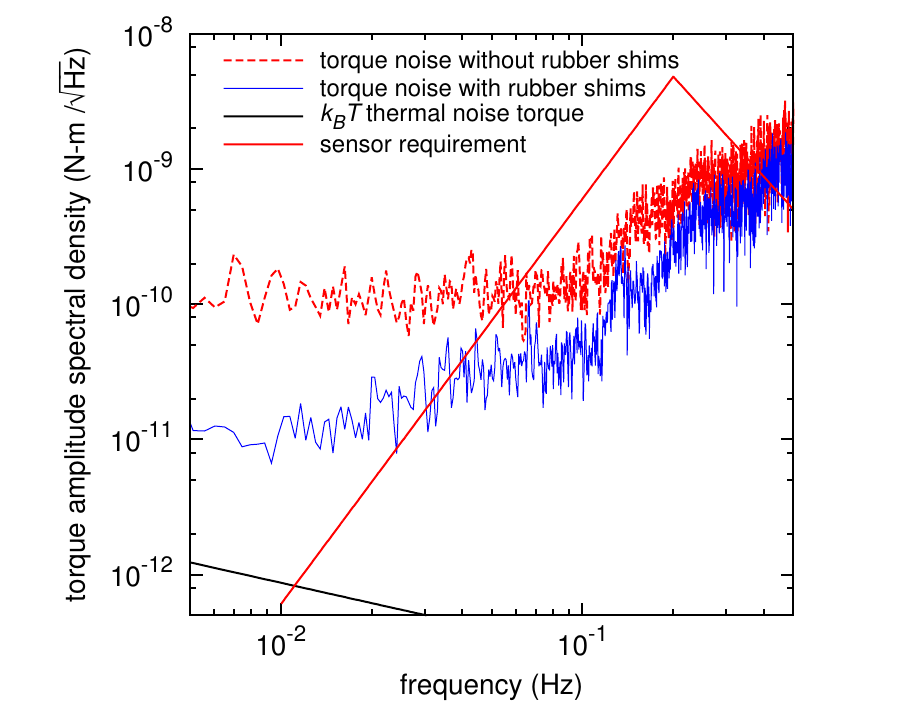}
\caption{\label{daily} Comparison of the lowest torque noise measured with and without the rubber shims placed under the feet of the support platform, demonstrating the suppression of the torque noise.}
\end{figure}

One of the largest noise sources we had to overcome was what we referred to as vibration noise. The spectral shape of the noise was white (in torque), but it had a daily amplitude variation of a factor of $\sim3$, correlated with ambient vibration levels. There are several models which could explain this kind of noise. One possible explanation is that this noise is a result of irregular torques arising from fluctuations of the parameters of the flexure (parametric down-conversion) triggered by excessive high-frequency vibrations.

Regardless of the exact mechanism for the process, we found that placing small rubber shims under the feet of the large aluminum platform to which the rotation sensor was bolted reduced this vibration noise (Fig.~\ref{daily}). The white-noise floor was lowered by a factor of 10 with the introduction of the rubber shims. Still, this may remain the current limiting noise source and might be worse in places with large vibrations. We plan to investigate and reduce this noise source further. 

An important noise source is the presence of a temperature gradient between the two arms of the balance. This armlength effect was analyzed by Speake \textit{et al.} \cite{speake1} Using their formulation and a measured upper limit on the temperature gradient, we estimated the upper limit of the angle noise from this effect to be less than $5\times10^{-10}$ rad$/\sqrt{\text{Hz}}$ at 10 mHz. 

Noise from ambient magnetic field fluctuations exerting a torque on the balance are expected to be small. Magnetic materials were avoided in the construction of the balance. The magnetic moment of the balance was measured to be $<$ $10^{-5}$ J/T. With ambient field fluctuations expected to be no larger than $3\times10^{-8}$ T$/\sqrt{\text{Hz}}$, the angle noise from magnetic field fluctuation would be less than $2\times10^{-10}$ rad$/\sqrt{\text{Hz}}$.

Another possible source of noise was from ambient gravity gradients. Our balance has a gravity gradient sensitivity of better than $0.03$ Eotvos$/\sqrt{\text{Hz}}$ or $3\times10^{-11}$ s$^{-2}/\sqrt{\text{Hz}}$ around $10$ mHz. For reference, this would allow the rotation sensor to resolve the gradient signal from a 20-ton bus, with closest approach of 80 m at $45^\circ$ to the beam axis, in $\sim100$ seconds to 2$\sigma$ under quiet conditions. However, the expected gradient noise due to environmental sources such as nearby trees, buildings, etc., was estimated from models to be lower than our current best sensitivity by at least an order of magnitude.

\begin{figure}
\centering
\includegraphics{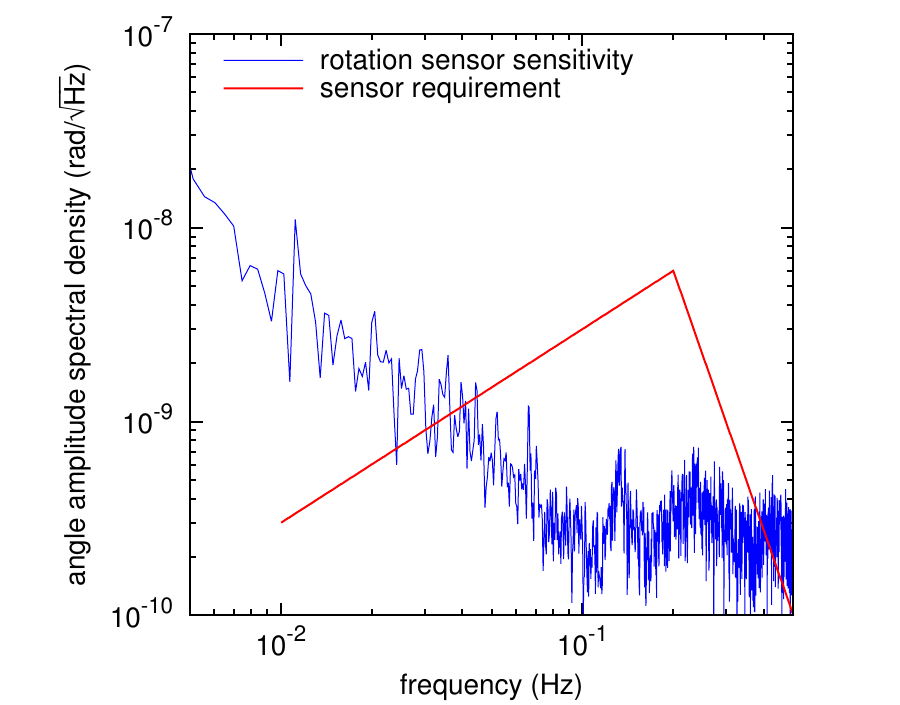}
\caption{\label{sens}Sensitivity of the rotation sensor.}
\end{figure}
\section{Discussion}
We have constructed a prototype absolute rotation sensor comprising of a $10.8$-mHz beam balance and an autocollimator. We also constructed a tiltmeter or inclinometer to measure background rotation at frequencies below $\sim60$ mHz. The two instruments show good correlation in the $15$-to $60$-mHz band during the day when background rotations are larger. Under quiet conditions the correlation is small, indicating that the instruments are limited by intrinsic noise. Using the best quiet-condition data, we were able to place upper limits on the sensitivity of the instrument as shown in Fig.~\ref{sens}. The sensor can reject horizontal acceleration to better than $3\times10^{-5}$ rad/m at frequencies above resonance. By design, the rotation sensor has optimal sensitivity in a frequency band spanning $\sim$few mHz to Hz. The low-frequency range of the sensor is limited by the $1/f^2$ frequency-dependence in sensitivity below the $10.8$ mHz resonance frequency of the balance. Above 0.1 Hz, the rotation sensor is limited by the $0.2$ nrad$/\sqrt{\text{Hz}}$ flat angular sensitivity of the autocollimator. At the high-frequency end, the sensor is limited by the 3.3 Hz downsampling of the autocollimator signal.If desired, it can be operated at frequencies as high as 1.5 kHz, excepting acoustic resonances of the balance.

This instrument has the potential to improve seismic isolation in Advanced LIGO by reducing the rotation noise contribution in horizontal seismometers. It meets the requirements described by Lantz \textit{et al.}\cite{reqs} above 40 mHz. We continue to improve the sensitivity of the instrument at low frequencies and to develop a more robust and compact version.

\begin{acknowledgments}
We would like to thank Ron Musgrave, Larry Stark and Jim Greenwell at the UW Physics Instrument shop for our flexures. We would like to thank the US taxpayers, NSF (grants: PHY0969488 and PHY1306613), and the LIGO Scientific Collaboration for funding and supporting this project. In particular, the authors would like to thank Vladimir Dergachev, Brian Lantz, Fabrice Matichard, and Rainer Weiss for valuable suggestions and discussions. We are also grateful to our colleagues in the Eot-Wash group for useful discussions. We thank the Center for Experimental Nuclear Physics and Astrophysics (CENPA) for use of its facilities.
\end{acknowledgments}


\end{document}